\documentclass[conference]{IEEEtran}
\IEEEoverridecommandlockouts
\usepackage{amsmath,amssymb,amsfonts}
\usepackage{cite}
\usepackage{algorithmic}
\usepackage{euscript,graphics}
\usepackage{textcomp}
\usepackage{xcolor}
\usepackage{times,amssymb,amsmath,amsfonts,float,nicefrac,color,subfigure}
\usepackage[ruled,vlined]{algorithm2e}
\usepackage{tikz}
\usepackage{bbding}
\usepackage{diagbox}
\usepackage{euscript,graphics}
\usepackage[dvips]{epsfig}
\usepackage{verbatim}
\usepackage{pgfplots}
\usepackage{multirow}
\usepackage{arydshln}
\usepackage{pgf}
\def\BibTeX{{\rm B\kern-.05em{\sc i\kern-.025em b}\kern-.08em
    T\kern-.1667em\lower.7ex\hbox{E}\kern-.125emX}}

\newtheorem{theorem}{Theorem}
\newenvironment{thmbis}[1]
  {\renewcommand{\thetheorem}{\ref{#1}$'$}%
   \addtocounter{theorem}{-1}%
   \begin{theorem}}
  {\end{theorem}}
\newtheorem{lemma}{Lemma}
\newtheorem{proposition}{Proposition}
\newtheorem{corollary}[proposition]{Corollary}

\newtheorem{remark}{Remark}  

\newcommand{\qbin}[3]{{#1 \brack #2}_{#3}}

\newcommand\nc\newcommand
\nc\bfa{{\boldsymbol a}}\nc\bfA{{\bf A}}\nc\cA{{\mathcal A}}
\nc\bfb{{\boldsymbol b}}\nc\bfB{{\bf B}}\nc\cB{{\mathcal B}}
\nc\bfc{{\boldsymbol c}}\nc\bfC{{\bf C}}\nc\cC{{\mathcal C}}
\nc\bfd{{\boldsymbol d}}\nc\bfD{{\bf D}}\nc\cD{{\mathcal D}}
\nc\bfe{{\boldsymbol e}}\nc\bfE{{\bf E}}\nc\cE{{\mathcal E}}
\nc\bff{{\boldsymbol f}}\nc\bfF{{\bf F}}\nc\cF{{\mathcal F}}
\nc\bfg{{\boldsymbol g}}\nc\bfG{{\bf G}}\nc\cG{{\mathcal G}}
\nc\bfh{{\boldsymbol h}}\nc\bfH{{\bf H}}\nc\cH{{\mathcal H}}
\nc\bfi{{\boldsymbol i}}\nc\bfI{{\bf I}}\nc\cI{{\mathcal I}}
\nc\bfj{{\boldsymbol j}}\nc\bfJ{{\bf J}}\nc\cJ{{\mathcal J}}
\nc\bfk{{\boldsymbol k}}\nc\bfK{{\bf K}}\nc\cK{{\mathcal K}}
\nc\bfl{{\boldsymbol l}}\nc\bfL{{\bf L}}\nc\cL{{\mathcal L}}
\nc\bfm{{\boldsymbol m}}\nc\bfM{{\bf M}}\nc\cM{{\mathcal M}}
\nc\bfn{{\boldsymbol n}}\nc\bfN{{\bf N}}\nc\cN{{\mathcal N}}
\nc\bfo{{\boldsymbol o}}\nc\bfO{{\bf O}}\nc\cO{{\mathcal O}}
\nc\bfp{{\boldsymbol p}}\nc\bfP{{\bf P}}\nc\cP{{\mathcal P}}
\nc\bfq{{\boldsymbol q}}\nc\bfQ{{\bf Q}}\nc\cQ{{\mathcal Q}}
\nc\bfr{{\boldsymbol r}}\nc\bfR{{\bf R}}\nc\cR{{\mathcal R}}
\nc\bfs{{\boldsymbol s}}\nc\bfS{{\bf S}}\nc\cS{{\mathcal S}}
\nc\bft{{\boldsymbol t}}\nc\bfT{{\bf T}}\nc\cT{{\mathcal T}}
\nc\bfu{{\boldsymbol u}}\nc\bfU{{\bf U}}\nc\cU{{\mathcal U}}
\nc\bfv{{\boldsymbol v}}\nc\bfV{{\bf V}}\nc\cV{{\mathcal V}}
\nc\bfw{{\boldsymbol w}}\nc\bfW{{\bf W}}\nc\cW{{\mathcal W}}
\nc\bfx{{\boldsymbol x}}\nc\bfX{{\bf X}}\nc\cX{{\mathcal X}}
\nc\bfy{{\boldsymbol y}}\nc\bfY{{\bf Y}}\nc\cY{{\mathcal Y}}
\nc\bfz{{\boldsymbol z}}\nc\bfZ{{\bf Z}}\nc\cZ{{\mathcal Z}}

\DeclareSymbolFont{bbold}{U}{bbold}{m}{n}
\DeclareSymbolFontAlphabet{\mathbbold}{bbold}

\begin{document}

\title{Some Results on the Improved Bound and Construction  of Optimal $(r,\delta)$  LRCs 
\thanks{
*Corresponding Author}
}

\author{\IEEEauthorblockN{Bin Chen\textsuperscript{1}, ~Weijun Fang\textsuperscript{2*}, ~Yueqi Chen\textsuperscript{3}, ~Shu-Tao Xia\textsuperscript{3}, ~Fang-Wei Fu\textsuperscript{4}, ~Xiangyu Chen\textsuperscript{5}}
\IEEEauthorblockA{\textit{ \textsuperscript{1} Harbin Institute of Technology, Shenzhen 518055, China.} \\
\textit{\textsuperscript{2}School of Cyber Science and Technology, Shandong University, Qingdao, Shandong, 266237, P.R. China.}\\
\textit{\textsuperscript{3}Tsinghua Shenzhen International Graduate School, Tsinghua University, Shenzhen 518055, China.}\\
\textit{\textsuperscript{4}Chern Institute of Mathematics and LPMC, Nankai University, Tianjin 300071, China.}\\
\textit{\textsuperscript{5}HUAWEI Technologies Co., Ltd., Shenzhen, 518055, P.R. China.}\\
 E-mail: chenbin2021@hit.edu.cn, nankaifwj@163.com, chen-yq20@mails.tsinghua.edu.cn, xiast@sz.tsinghua.edu.cn,\\ fwfu@nankai.edu.cn, chenxiangyu9@huawei.com.}
}

\maketitle

\begin{abstract}
Locally repairable codes (LRCs) with $(r,\delta)$ locality were introduced by Prakash \emph{et al.} into distributed storage systems (DSSs) due to their
benefit of locally repairing at least $\delta-1$ erasures via other $r$ survival nodes among the same local group.  An LRC achieving the $(r,\delta)$ Singleton-type bound is called an optimal $(r,\delta)$ LRC. Constructions of optimal $(r,\delta)$ LRCs with longer code length and determining the maximal code length have been an important research direction in coding theory in recent years. In this paper, we conduct further research on the improvement of  maximum code length of optimal $(r,\delta)$  LRCs. For $2\delta+1\leq d\leq 2\delta+2$, our upper bounds largely improve the ones by Cai \emph{et al.}, which are tight in some special cases. Moreover, we generalize the results of Chen \emph{et al.} and obtain a complete characterization of optimal $(r=2, \delta)$-LRCs in the sense of geometrical existence in the finite projective plane $PG(2,q)$. Within this geometrical characterization, we construct a class of optimal $(r,\delta)$ LRCs based on the sunflower structure. Both the construction and upper bounds are better than previous ones.
\end{abstract}


\section{Introduction}
In the era of big data, the reliability
of large-scale data storage becomes more and more important. Distributed storage system (DSS) with erasure codes provides a new paradigm to maintain high reliability and flexibility. Regenerating codes \cite{dimakis2010network} and locally repairable codes (LRCs) \cite{GHSY12} are the two most representative erasure coding schemes in a DSS, which respectively optimize the repair bandwidth and repair locality. We mainly focus on LRCs in this paper, which have been implemented and evaluated in two DSSs, Windows Azure storage \cite{windows} and Hadoop Distributed File System \cite{hadoop}. Next, we formally give the definition of LRCs in \cite{GHSY12}. Let $\mathcal{C}$ be a $q$-ary $[n, k, d]$ linear code with dimension  $k$ and minimum distance $d$. We say that a linear code $\mathcal{C}$ is an $r$-local LRC with \emph{locality} $r$ if for any $i\in[n]\triangleq\{1, 2,\dots, n\}$, there exists a subset $R_i\subset [n]\setminus \{i\}$ with $|R_{i}| \leq r$ such that the $i$-th symbol $c_i$ can be recovered by $\{c_j\}_{j\in R_i}$, i.e., $c_i$ can be represented as a linear combination of $\{c_j\}_{j\in R_i}$. In \cite{GHSY12}, the authors also derived the well-known Singleton-type bound on $d$ for an $r$-local $q$-ary $[n, k, d]$-LRCs, 
\begin{equation}
\label{singleton}
d\leq n-k-\left\lceil \frac{k}{r}\right\rceil+2.
\end{equation}
Note that bound \eqref{singleton} reduces to the classical Singleton bound when $r=k$. Optimal $r$-local LRCs  achieving the bound \eqref{singleton} with equality have been obtained in recent years \cite{TB14,tamo2016optimal,silberstein2013optimal,LXY19,CXHF18,CFXF19,J19,li2018optimal,xing2018construction}. For general $d$ and $r$, optimal Reed-Solomon-like LRCs with lengths $n\leq q-1$ \cite{TB14} are obtained based on the polynomial evaluation. The reference \cite{tamo2016optimal} 
establishes a connection between the construction of optimal LRCs and properties of the matroid represented by its generator matrix. A novel explicit construction of LRCs via maximum rank distance codes is obtained in \cite{silberstein2013optimal}, which provides new optimal vector and scalar LRCs. Optimal cyclic or constacyclic LRCs are obtained in \cite{CXHF18, CFXF19, LXY19}. \cite{J19} presents an optimal explicit construction of LRCs with distance $d=5$ and 6 via binary constant weight codes. For some particular $r=2,3,5,7,11, 23$, optimal LRCs with lengths up to $q+2\sqrt{q}$ are constructed via elliptic curves \cite{li2018optimal}. In addition to the progress, the study of maximal code length \cite{GXY19, CFXH19} and optimal $r$-local LRCs with longer code length \cite{CXHF18,CFXF19,J19,li2018optimal,xing2018construction} has important value in theory and practice, especially on the design of LRCs with flexible parameters since shorter optimal LRCs can be obtained from the longer ones. 


To tolerate multiple erasures in each local group, LRCs with $(r,\delta)$ locality were introduced by Prakash \emph{et al} \cite{prakash2012optimal}. Other generalizations include code with hierarchical locality \cite{sasidharan2015codes}, or with 
availability \cite{wang2014repair,rawat2016locality}, etc. Specifically, the $i$-th symbol $c_i$ of a $q$-ary $[n, k]$ linear code $\cC$ is said to have $(r, \delta)$-locality ($\delta\ge2$) if  there exists a punctured subcode of $\cC$ with support containing $i$, whose length is at most $r + \delta - 1$, and whose minimum distance is at least $\delta$, i.e., there exists a subset $S_{i}\subseteq [n]\triangleq\{1,2,\ldots,n\}$ such that $i\in S_{i}$, $|S_{i}|\le r+\delta-1$ and $d(\cC|_{S_{i}})\ge\delta$. A generalized Singleton-type bound on $d$ was also derived as follows. 
\begin{equation} \label{GeneralizedSingleton}
d \leq n-k+1-\left( \left \lceil \frac{k}{r} \right \rceil-1 \right)(\delta -1).
\end{equation}
Following the research line of $r$-local LRCs, various generalized constructions of optimal $(r,\delta)$ LRCs achieving the bound \eqref{GeneralizedSingleton} were obtained in \cite{Cai, cai2, zhang20, kong}. When $\delta+1\leq d\leq 2\delta$, optimal $(r,\delta)$ LRCs with unbounded length were given in \cite{fang2020optimal, zhang20}. For $d\geq 2\delta+1$, Cai \emph{et al} \cite{Cai} first obtain upper bounds on the code length and constructions of optimal $(r,\delta)$ LRCs, which generalize previously ones by Guruswami \emph{et al} \cite{GXY19} under weaker conditions. For $d\geq 3\delta+1$ and $r\geq d-\delta$, the authors \cite{kong}
have generalized the results of \cite{xing2018construction} based on the sparse hypergraphs to obtain optimal $(r,\delta)$ LRCs with super-linear in the alphabet size. To the best of our knowledge, known results on the maximal code length of $(r,\delta)$ LRCs is not tight in general, and the constructions of $(r,\delta)$ LRCs with longer code length only focus on the range of larger localities that are less attractive in a practical DSS. 

In this paper, we instead focus on the maximum code length of $q$-ary optimal $(r,\delta)$ LRCs with small localities. To simplify the discussion, we assume that an $(r,\delta)$ LRC has \emph{disjoint local subcodes}, i.e, $(r+\delta-1)\mid n$ and each subcode $\cC|_{S_{i}}$ is an $[r+\delta-1, r, \delta]$ MDS code. By generalizing the parity-check matrix approach (\cite{Haoarxiv, CFXH19}) for $r$-local LRCs, we can similarly obtain the \emph{standard-form} parity-check matrix for an  $(r,\delta)$ LRCs with disjoint local subcodes. Within this standard-form parity-check matrix, we can obtain an improved upper bound on the number $\ell$ of disjoint local subcodes via a novel linear combination among the columns of each local subcode. Thus we derive an improved upper bound on the code length compared with that of Cai \emph{et al}. In addition, we can show that our new bounds are exactly achieved by some known constructions. Based on the complete characterization of optimal $r$-local LRCs with $d=6$ and $r=2$ \cite{CFXH19}, we establish a generalized characterization of optimal $(r=2,\delta)$ LRCs with $d=2\delta+2$. This geometrical characterization further help us to verify the existence of optimal $(r=2,\delta)$ LRCs with $n=(\delta+1)(q+1)$ and $d=2\delta+2$ via the sunflower structure, which forms a generalization of an  optimal quaternary $(r,\delta)$ LRC in \cite{shumfour}. Following the line-point incidence matrix approach \cite{CFXH19}, we further obtain three improved upper bounds on the maximal code length based on the Johnson bound for constant weight codes. To the best of our knowledge, our bounds are the best with the lowest order of $q$ for optimal $(r=2,\delta)$ LRCs with $d=2\delta+2$ until now.

The rest of this paper is organized as follows. In Section II, we introduce some necessary results on the standard-form parity-check matrix and the Johnson upper bound for binary constant weight codes. In Section III, we establish a generalized complete characterization for optimal $(r=2,\delta)$ LRCs with $d=2\delta+2$ and present a class of optimal $(r=2,\delta)$ LRCs via the sunflower structure. Furthermore, we derive some improved upper bounds  based on the incidence matrix and Johnson bound. Section IV concludes this paper.

\section{Preliminaries}
In this section, we introduce the basic parity matrix structure of optimal $(r,\delta)$ LRCs with disjoint local groups, and the Johnson bound for binary constant weight codes that are utilized to obtain our improved upper bounds on the maximal code length. Let $\mathbb{F}_q$ be a finite field with $q$ elements, $\mathbb{F}_q^{*}$ denotes the set of its nonzero elements. Throughout this paper, we assume that $(r+\delta-1) \mid n$ and the LRCs have \emph{disjoint local repair groups}.  By the definition of $(r,\delta)$ locality \cite{prakash2012optimal}, we know that an $[n, k, d]$-linear code $C$ has $(r,\delta)$ locality if there exist $\ell\triangleq \frac{n}{r+\delta-1}$ punctured $[r+\delta-1, r, \delta]$ MDS subcodes, i.e, there exist $\ell$ submatrices $\mathbf{H}_{1}, \mathbf{H}_{2}, \ldots, \mathbf{H}_{\ell}\in \mathbb{F}_q^{(\delta-1)\times (r+\delta-1)}$ which are called \emph{locality matrices}, such that they respectively form a parity-check matrix of the $[r+\delta-1, r, \delta]$ MDS subcode. Thus an optimal $(r,\delta)$ LRC with disjoint local repair groups has an equivalent parity-check matrix $\mathbf{H}$ as the following form:
 \begin{equation}\label{pcm}
 \setlength{\arraycolsep}{2.3pt}
\mathbf{H}\!=\!\!\left(
  \begin{array}{cc|cc|c|cc}
    \mathbf{I}_{\delta-1} & \mathbf{Q}_1 &    &   &    &    &  \\
     &   &    \mathbf{I}_{\delta-1} & \mathbf{Q}_2   &    &   &  \\
     &   &   &   & \ddots   &   &    \\
       &   &     &      &    & \mathbf{I}_{\delta-1} & \mathbf{Q}_{\ell}   \\
    \hline
    \mathbf{0}_{u\times (\delta-1)} &  \mathbf{V}_{1} & \mathbf{0}_{u\times (\delta-1)}   &   \mathbf{V}_{2} & \dots & \mathbf{0}_{u\times (\delta-1)}    & \mathbf{V}_{\ell}
  \end{array}
\right),\!\!
\end{equation}
where the upper part of $\mathbf{H}$ contains $\ell$ locality subblocks and the lower part of $\mathbf{H}$ contains $u\triangleq n-k-\ell(\delta-1)$ rows which ensure that $rank(H)=n-k$. Note that each submatrix $\mathbf{H}_i=( \mathbf{I}_{\delta-1} \mathbf{Q}_i )$ exactly forms a standard generator matrix of a $[r+\delta-1, \delta-1, r+1]$ MDS code, where $\mathbf{I}_{\delta-1}$ is an identity matrix with unit column vector $\mathbf{e}_s\in \mathbb{F}_q^{(\delta-1)}$, $s\in[\delta-1]$, and each element of $\mathbf{Q}_i, i\in[\ell]$ is nonzero, i.e., $\mathbf{Q}_i\in{\mathbb{F}_q^{*}}^{{(\delta-1)}\times r}$. The bold-type letters $\mathbf{V}_{i}$, $i\in[\ell]$, are submatrices in $\mathbb{F}_q^{{u}\times (\delta-1)}$, while $\mathbf{0}_{u\times (\delta-1)}  $ represents the all-zero matrix in $\mathbb{F}_q^{{u}\times (\delta-1)}$.

Next, we recall Lemma II.2 in \cite{kong}, which relates the number of rows in the low part of $H$ with other code parameters. 
\begin{lemma}(\cite{kong})\label{lemma1}
Let $n,k,d,r,\delta$ be positive integers with $(r+\delta-1)\mid n$. If the generalized Singleton-type bound \eqref{GeneralizedSingleton} is achieved, then
\begin{equation}\label{eq:2}
n-k=\ell(\delta-1)+d-1-\left(\left\lfloor\frac{d-\delta}{r+\delta-1}\right\rfloor+1\right)(\delta-1),
\end{equation}
\begin{equation} 
i.e,\,u=d-1-\left(\left\lfloor\frac{d-\delta}{r+\delta-1}\right\rfloor+1\right)(\delta-1).
\end{equation}
\end{lemma}

We denote by $(n, M, d; w)$ a binary constant-weight code of length $n$, size $M$, minimum
distance $d$, and each codeword has a fixed Hamming weight $w$. The well-known Johnson bound can be summarized as follows.
\begin{lemma}(\cite[Johnson Bound]{MacWilliams})\label{prop1}
Let $C$ be a binary $(n, M, d=2\delta; w)$-constant weight code, then we have 
\begin{eqnarray}\label{ujohnson}
A(n, 2\delta, w)&\leq&\frac{\binom{n}{w-\delta+1}}{\binom{w}{w-\delta+1}}.\\
M(w^2-wn+\delta n)&\leq& \delta n.
\end{eqnarray}
\end{lemma}

\section{Main Results}
In this section, we always assume that an $(r,\delta)$ LRC has disjoint local repair groups, and denote
\begin{eqnarray*}
n_{\max}(q,d,r,\delta)&\!\!\!=\!\!\!&\max\{n\mid \text{there exists a $q$-ary}\\ &\!\!\!\!\!\!&\qquad\qquad\text{optimal $(n, k, d;r,\delta)$ LRC}\}.
\end{eqnarray*}

\subsection{General Improved Upper bound}
\begin{lemma}\label{klemma} Let $H$ be an $(\ell(\delta-1)+u) \times$ $\ell(r+\delta-1)$ parity-check matrix of an $(r,\delta)$ LRC with disjoint local repair groups. If $d \geq 2\delta+1$, then
\begin{equation} 
\ell \leq \left\lfloor (q^{u}-1)/(q-1) \cdot\left(\begin{array}{c}
r+\delta-1\\
\delta
\end{array}\right)\right\rfloor
\end{equation}
\end{lemma}

\begin{IEEEproof}
Since $d\geq 2\delta+1$, we know that each column of $\mathbf{V}_i$, $i\in[\ell]$, is a nonzero vector in  $\mathbb{F}_q^{u}$. For each local group, we have $\binom{r+\delta-1}{\delta}$ choices of selecting $\delta$ columns. Note that each $\mathbf{H}_i$ is an $[r+\delta-1, r, \delta]$ MDS code, any $\delta$ columns of $\mathbf{H}_i$ must be linearly dependent, i.e., there exists some nonzero linear combination to make these $\delta$ columns contribute to the zero vector. Accordingly, the nonzero linear combination also produces a nonzero vector in $\mathbb{F}_q^{u}$ at the lower part. In total, we can produce $\ell\cdot\binom{r+\delta-1}{\delta}$ distinct nonzero vectors in $\mathbb{F}_q^{u}$ because $d \geq 2\delta+1$. Then we have 
$\ell \leq \left\lfloor (q^{u}-1)/(q-1) \cdot\left(\begin{array}{c}
r+\delta-1\\
\delta
\end{array}\right)\right\rfloor$.
\end{IEEEproof}

By Lemma\ref{klemma}, we can directly obtain the following improved upper bounds on the code length.
\begin{theorem}\label{thm1}
Let $\cC$ be a Singleton-optimal $(n, k, d;r, \delta)$ LRC with disjoint local repair groups.
\begin{enumerate}
\item[(i)]If $d=2\delta+1$, then
\begin{equation}\label{bound1}
 n  \le  (r+\delta-1)\left\lfloor\frac{q^{u}-1}{(q-1) \cdot\left(\begin{array}{c}
r+\delta-1\\
\delta
\end{array}\right)}\right\rfloor
\end{equation}
where $u=\delta+1-\left(\left\lfloor\frac{\delta+1}{r+\delta-1}\right\rfloor+1\right)(\delta-1)$
\item[(ii)]If $d=2\delta+2$, then
\begin{equation}\label{bound2}
  n\le  (r+\delta-1)\left\lfloor\frac{q^{u}-1}{(q-1) \cdot\left(\begin{array}{c}
r+\delta-1\\
\delta
\end{array}\right)}\right\rfloor,
\end{equation}
where $u=\delta+2-\left(\left\lfloor\frac{\delta+2}{r+\delta-1}\right\rfloor+1\right)(\delta-1)$
\item[(iii)]If $d=3\delta$, then
\begin{equation}\label{bound2}
  n\le  (r+\delta-1)\left\lfloor\frac{q^{u}-1}{(q-1) \cdot\left(\begin{array}{c}
r+\delta-1\\
\delta
\end{array}\right)}\right\rfloor,
\end{equation}
where $u=2\delta-\left(\left\lfloor\frac{2\delta}{r+\delta-1}\right\rfloor+1\right)(\delta-1)$
\end{enumerate}
\end{theorem}

\begin{remark}
Note that 
$$2\delta+1\leq d\leq 3\delta\Leftrightarrow 2\leq \frac{d-1}{\delta}<3\Leftrightarrow t\triangleq\lfloor\frac{d-1}{\delta}\rfloor=2,$$
we can directly write out the upper bound of \cite[Theorem III.3 ]{Cai} for $2\delta+1\leq d\leq 3\delta$:
\begin{equation}\label{cai}
  n \le (r+\delta-1)\left\lfloor\frac{q^u}{r(q-1)}\right\rfloor.
\end{equation}
\end{remark}

Clearly, our bounds in Theorem \ref{thm1} are always tighter than the bound (\ref{cai}). Moreover,  our upper bounds (\ref{bound1})-(\ref{bound2}) in Theorem \ref{thm1} can be exactly achieved as shown below.

\begin{corollary}Let $\cC$ be a Singleton-optimal $(n, k, d;r, \delta)$ LRC with disjoint local repair groups.
\begin{enumerate}
\item[(i)]If $d=2\delta+1$ and $r=2$, then
\begin{equation}\label{q+1}
  n \le q+1.
\end{equation}
Moreover, $n_{\max}(q,2\delta+1,2,\delta)=q+1$.
\item[(ii)]If $d=2\delta+2$ and $r=2$, then
\begin{equation}\label{ternary}
 n \le (\delta+1)\cdot \left\lfloor\frac{q^2+q+1}{\delta+1}\right\rfloor.
\end{equation}
Moreover, $n_{\max}(4,8,2,3)=20$.
\end{enumerate}
\end{corollary}
\begin{IEEEproof}
\begin{enumerate}
\item[(i)]
The upper bound \eqref{q+1} can be easily derived by \eqref{bound1}. And the constructions of optimal cyclic and constacyclic $(r,\delta)$ LRCs \cite{CXHF18,CFXF19} implies
$n_{\max}(q,2\delta+1,2,\delta)\geq q+1$. Therefore, $n_{\max}(q,2\delta+1,2,\delta)=q+1$.
\item[(ii)] We can directly obtain the upper bound by \eqref{bound2}. On the other hand, the existence of optimal quaternary $(r=3, \delta=3)$ LRC \cite[(36)]{shumfour} implies that $n_{\max}(4,8,2,3)\geq 20$. Therefore, $n_{\max}(4,8,2,3)=20$.
\end{enumerate}
We finish the proof. 
\end{IEEEproof}

\subsection{Complete Characterization of Optimal $(n,k,d=2\delta+2;r=2,\delta)$-LRCs}
In the following parts, we suppose $r=2$, $(\delta+1) \mid n$ and $\ell \triangleq n/(\delta+1)$. Let $C$ be an optimal $(n, k, d=2\delta+2; r=2, \delta)$-LRC with disjoint local repair groups, then $u\triangleq n-k-\ell(\delta-1)=3$ by Lemma \ref{lemma1}.
By \eqref{pcm}, we may assume that $C$ has a parity-check matrix $H$ as follows:
\begin{equation}\label{2.1}
\left(
  \begin{array}{ccc|ccc|c|ccc}
    \mathbf{I}_{\delta-1} & \mathbf{p}_1 &  \mathbf{q}_1 & 0 & 0 &  0 & \ldots & 0 & 0 & 0 \\
    0 & 0 & 0 & \mathbf{I}_{\delta-1} & \mathbf{p}_2 &  \mathbf{q}_2 & \ldots & 0 & 0 & 0 \\
    \vdots & \vdots & \vdots & \vdots &  \vdots & \vdots & \ddots & \vdots & \vdots & \vdots \\
    0 & 0 &  0 & 0 & 0 & 0 & \ldots & \mathbf{I}_{\delta-1} & \mathbf{p}_{\ell} &  \mathbf{q}_{\ell} \\
    \hline
    \textbf{0} & \textbf{u}_{1} &  \textbf{v}_{1} & \textbf{0} & \textbf{u}_{2} & \textbf{v}_{2} & \ldots & \textbf{0} & \textbf{u}_{\ell} & \textbf{v}_{\ell} \\
  \end{array}
\right),
\end{equation}
where $\textbf{u}_{i},\textbf{v}_{i} \in \mathbb{F}_{q}^{3},$ $i \in [\ell]$.
Denote $\mathcal{V}_{i}=\textnormal{Span}\{\textbf{u}_{i},\textbf{v}_{i}\}$ to be a subspace of $\mathbb{F}_{q}^{3}$ spanned by $\textbf{u}_{i}$ and $\textbf{v}_{i}$ over $\mathbb{F}_{q}$. Since each submatrix $(\mathbf{I}_{\delta-1}, \mathbf{p_i}, \mathbf{q_i})$ forms a standard generator matrix of an MDS code,  we know that $\mathbf{p}_i\triangleq(p_{i,1}, p_{i,2}, \cdots, p_{i,(\delta-1)})^{T}, \mathbf{q}_i\triangleq(q_{i,1}, q_{i,2}, \cdots, q_{i,(\delta-1)})^{T}\in {\mathbb{F}_q^{*}}^{(\delta-1)}, i\in[\ell]$ by the property of MDS codes \cite{MacWilliams}. Thus for any fixed $m\in [\delta-1]$, there must exist $(a_m, b_m)\in \mathbb{F}_q^{*}\times \mathbb{F}_q^{*}$ such that 
\begin{equation}\label{ex}
a_mp_{i,m}+b_mq_{i,m}=0. 
\end{equation}
We let $\mathcal{F}_i\subseteq \mathbb{F}_q^{*}\times \mathbb{F}_q^{*}, i\in[\ell]$ to be the set containing all of these tuples $(a_m,b_m)$ for the $i$-th repair group. Then we can obtain the following lemma to prove our complete  characterization.
\begin{lemma}\label{lem3}
  Suppose $q\ge \delta+1$ and $(\delta+1) \mid n$. Let $C$ be a $q$-ary optimal $(n,k, d=2\delta+2; r=2, \delta)$-LRC with a parity-check matrix $H$ as (\ref{2.1}). Then
\begin{itemize}
  \item[(i)]  $\dim(\mathcal{V}_{i})=2;$
  \item[(ii)] for any $j \neq i \in [\ell]$, we have $\textbf{u}_{i},\textbf{v}_{i}, a_m\textbf{u}_{i}+b_m\textbf{v}_{i} \notin \mathcal{V}_{j}$, where $(a_m, b_m)\in \mathcal{F}_i$, $m\in[\delta-1]$.
\end{itemize}
\end{lemma}

\begin{IEEEproof}
\begin{enumerate}
\item[(i)] If $\dim(\mathcal{V}_{i}) \leq 1$ for some $i$, then the $\delta+1$ columns in the $i$-th repair group of $H$ are linearly dependent, which leads to $d \leq \delta+1$. This contradicts with $d=2\delta+2$.

\item[(ii)] Suppose for some $j \neq i \in [\ell]$, one of $\textbf{u}_{i},\textbf{v}_{i}$ belongs to $\mathcal{V}_{j}$. Without of loss generality, we assume $i=1$ and $ j=2$. If $\textbf{u}_{1} \in \mathcal{V}_{2}$, then $\textbf{u}_{1}=a\textbf{u}_{2}+b\textbf{v}_{2}$ for some $a, b \in \mathbb{F}_{q}$. Then it is easy to verify that the first $\delta$ columns in the first repair group and the $\delta+1$ columns in the second repair group of $H$  are linearly dependent, which leads to $d \leq 2\delta+1$ and contradicts with the fact that $d= 2\delta+2$. Similarly, we can prove $\textbf{v}_{1} \notin \mathcal{V}_{2}$. 

Finally, we assume that for some $m_0\in [\delta-1]$, there exists  $(a_{m_0}, b_{m_0})\in \mathcal{F}_i$ such that $a_{m_0}\textbf{u}_{1}+b_{m_0}\textbf{v}_{1} \in \mathcal{V}_{2}$. Then we can show that all columns but the $m_0$-th one in the first repair group, and the  $\delta+1$ columns in the second repair group are linearly dependent, which also leads to $d \leq 2\delta+1$. 
\end{enumerate}
The proof is completed.
\end{IEEEproof}

For each 2-subspace $\mathcal{V}_{i}$  of $\mathbb{F}_{q}^{3}$, $i\in[\ell]$, we define $\qbin{\mathcal{V}_{i}}{1}{q}$ used in \cite{Raviv} to be the set of all 1-subspaces of $\mathcal{V}_{i}$. Clearly, $\left|\qbin{\mathcal{V}_{i}}{1}{q}\right|=q+1$. To avoid the degenerated case when $\delta=2$ as in \cite[Theorem]{MacWilliams}, we assume that $\delta>2$, i.e, $\delta-1\geq 2$ below. Then we can obtain the following claim.

\emph{Claim 1:} For $m, m'\in [\delta-1]$, $m\neq m'$, the generated 1-subspaces $\textnormal{Span}\{a_m\textbf{u}_{i}+b_m\textbf{v}_{i}\}$  and $\textnormal{Span}\{a_{m'}\textbf{u}_{i}+b_{m'}\textbf{v}_{i}\}$  in $\qbin{\mathcal{V}_{i}}{1}{q}$ are distinct. 
\begin{IEEEproof}
If $\textnormal{Span}\{a_m\textbf{u}_{i}+b_m\textbf{v}_{i}\}$  and $\textnormal{Span}\{a_{m'}\textbf{u}_{i}+b_{m'}\textbf{v}_{i}\}$ are the same 1-subspace, then there exists $c\in \mathbb{F}_q^{*}$ such that 
$$a_m\textbf{u}_{i}+b_m\textbf{v}_{i}=c(a_{m'}\textbf{u}_{i}+b_{m'}\textbf{v}_{i})\Leftrightarrow a_m=ca_{m'}, b_m=cb_{m'},$$
then we can show that the $(\delta-1)$ columns $\{\mathbf{e}_s\mid s\in[\delta-1]\setminus\{m,m'\}\}\cup\{\mathbf{p}_i, \mathbf{q}_i\}$ are linearly dependent, which contradicts with the fact that $(\mathbf{I}_{\delta-1}, \mathbf{p_i}, \mathbf{q_i})$ forms a standard generator matrix of a MDS code.
\end{IEEEproof}

Based on Claim 1 and the previous lemma, we can now obtain a complete characterization of optimal $(n,k,d=2\delta+2;r=2,\delta)$-LRCs as follows.
\begin{theorem}\label{thm2}
Suppose $q\ge \delta+1$ and $\delta+1 \mid n$. There exists a $q$-ary optimal $(n,k, d=2\delta+2; r=2, \delta)$-LRC with disjoint repair groups if and only if
there exist $\ell$ distinct 2-subspaces $\mathcal{V}_{1}, \mathcal{V}_{2}, \ldots, \mathcal{V}_{\ell}$ of $\mathbb{F}_{q}^{3}$  such that for each $i \in [\ell]$,
\[t_{i}\triangleq \left|\bigcup_{j\in[\ell],j\neq i}\left(\qbin{\mathcal{V}_{i}}{1}{q}\bigcap \qbin{\mathcal{V}_{j}}{1}{q}\right)\right|\le q-\delta.\]
\end{theorem}

\begin{IEEEproof}
The proof is divided into two directions as follows.
\begin{itemize}
\item \textbf{Necessity:} Let $C$ be an optimal $(n, k, d=2\delta+2; r=2, \delta)$-LRC with a parity-check matrix $H$. By Lemma \ref{lem3} and Claim 1, we have $\dim(\mathcal{V}_i)=2$ and $\textnormal{Span}\{\textbf{u}_{i}\}$, $\textnormal{Span}\{\textbf{v}_{i}\}$, $\textnormal{Span}\{a_m\textbf{u}_{i}+b_m\textbf{v}_{i}\}, m\in[\delta-1]$, are $(\delta+1)$ distinct 1-subspaces in $\qbin{\mathcal{V}_{i}}{1}{q}$, which do not belong to $\qbin{\mathcal{V}_{j}}{1}{q}$ for all $j\neq i$. Thus $t_{i} \leq q+1-(\delta+1)=q-\delta$.
\smallskip

\item \textbf{Sufficiency:} Suppose there exist $\ell$ distinct 2-subspaces $\mathcal{V}_{1}, \mathcal{V}_{2}, \ldots, \mathcal{V}_{\ell}$ of $\mathbb{F}_{q}^{3}$ satisfying the intersection condition in the theorem. Since  $t_{i} \leq q-\delta$, there exist $(\delta+1)$ distinct 1-subspaces of $\mathcal{V}_i$, without loss of generality, we let these  $(\delta+1)$ distinct 1-subspaces to be $\textnormal{Span}\{\textbf{u}_{i}\}$, $\textnormal{Span}\{\textbf{v}_{i}\}$, $\textnormal{Span}\{a_m\textbf{u}_{i}+b_m\textbf{v}_{i}\}, m\in[\delta-1]$, $(a_m, b_m)\in \mathbb{F}_q^{*}\times \mathbb{F}_q^{*}$. Similar to \eqref{ex}, we can show that there exists $(p_{i,m}, q_{i,m})\in \mathbb{F}_q^{*}\times \mathbb{F}_q^{*}$ such that
\begin{equation}\label{ffe}
a_mp_{i,m}+b_mq_{i,m}=0.
\end{equation}
Next, we need to prove that the matrix $H_i=(\mathbf{I}_{\delta-1}, \mathbf{p_i}, \mathbf{q_i})$  formed by these $(p_{i,m}, q_{i,m})\in \mathbb{F}_q^{*}\times \mathbb{F}_q^{*}$ is a standard generator matrix of a $[\delta+1, \delta-1, 3]$ MDS code, i.e, any $(\delta-1)$ columns of $H_i$ are linearly independent. Based on the structure of $H_i$, we only need to prove the last $(\delta-1)$ columns are linearly independent in the sense of equivalence. By contradiction, we assume that last $(\delta-1)$ columns are linearly dependent, then there must exist
$(a, b)\in \mathbb{F}_q^{*}\times \mathbb{F}_q^{*}$
such that $ap_{i,1}+bq_{i,1}=0$, $ap_{i,2}+bq_{i,2}=0$, i.e., $a=ca_1, b=cb_1$, $a=c'a_2, b=c'b_2$ for some $c,c'\in \mathbb{F}_q^{*}$. Thus $a_1=c^{-1}c'a_2$, $b_1=c^{-1}c'b_2$, which implies $\textnormal{Span}\{a_1\textbf{u}_{i}+b_1\textbf{v}_{i}\}$ and $\textnormal{Span}\{a_2\textbf{u}_{i}+b_2\textbf{v}_{i}\}$ contributes to the same 1-subspace. This makes a contradiction. Therefore, each matrix $H_i=(\mathbf{I}_{\delta-1}, \mathbf{p_i}, \mathbf{q_i})$ is indeed a standard generator matrix of a $[\delta+1, \delta-1, 3]$ MDS code, then a linear code $\cC$ with parity-check $H$ given as \eqref{2.1} has $(r=2,\delta)$ locality and dimension $k \geq n-\ell(\delta-1)-3=2\ell-3$. By \eqref{GeneralizedSingleton}, we have $d \leq 2\delta+2$. We only need to show that $d \geq 2\delta+2$. By contradiction, we assume that $d \leq 2\delta+1$, i.e., there exist $2\delta+1$ columns of $H$ which are linearly dependent. Due to the structure of $H$, we only need to prove that $\delta$
columns from one local group and  $\delta+1$ columns from another local group are linearly independent. Otherwise, we assume these $2\delta+1$ columns are linearly dependent and distributed in the first two repair local groups, then we know that $\textnormal{Span}\{\textbf{u}_{1}\}$, $\textnormal{Span}\{\textbf{v}_{1}\}$, or $\textnormal{Span}\{a_m\textbf{u}_{1}+b_m\textbf{v}_{1}\}, m\in[\delta-1]$ must belong to 
$\qbin{\mathcal{V}_{2}}{1}{q}$ by \eqref{ffe} and the structure of $H$, which leads to a contradiction. 
\end{itemize}
Thus we finish the proof.
\end{IEEEproof}

Let $PG(2,q)$ be the projective plane over finite field $\mathbb{F}_{q}$. We can restate Theorem \ref{thm2} in an equivalent geometrical language as \cite[Theorem 5']{CFXH19}.


\begin{thmbis}{thm2}[{\bf Geometrical Characterization of Optimal $(n,k,d=2\delta+2;r=2,\delta)$-LRCs}]\label{thm1.1}
  Suppose $q\ge \delta+1$ and $(\delta+1) \mid n$. Then, there exists an optimal $(n, k, d=2\delta+2, r=2, \delta)$-LRC with disjoint repair groups if and only if there exist $\ell$ distinct lines $L_{1},L_{2},\ldots,L_{\ell}$ in $PG(2,q)$, such that each $L_{i}$ has at most $(q-\delta)$ distinct intersection points.
\end{thmbis}

We refer to the condition in Theorem \ref{thm1.1} as \emph{intersection condition} in the sequel. From Theorem \ref{thm1.1}, we can employ the theory of finite geometry to study the construction and bound of optimal $(r,\delta)$ LRCs with minimum distance $d=2\delta+2$ and locality $r=2$. 
Note that the \emph{sunflower} structure \cite{Etzion} of $PG(2,q)$ naturally satisfies the intersection condition in Theorem \ref{thm1.1} when $q\geq \delta+1$ since all lines intersect in a common point. Then we can generalize the $r$-local sunflower construction \cite[Theorem 4]{CFXH19} to the $(r,\delta)$ case below. 

\begin{theorem}[{\bf $(r,\delta)$ Sunflower Construction}]\label{sc}
Let $q \geq \delta+1$ be a prime power, then there exists a $q$-ary optimal $(n=(\delta+1)(q+1),k=2q-1, d=2\delta+2 ;r=2,\delta)$-LRC.
\end{theorem}
\begin{remark}
Note that when $q=4$ and $\delta=3$, the construction (36) of \cite{shumfour} exactly corresponds to the sunflower constructions  and achieves the maximal code length. 
\end{remark}

\subsection{Improved Upper Bounds for optimal $(n,k,d=2\delta+2;r=2,\delta)$ LRCs based on Constant Weight Codes}
Similar to \cite[Theorem 9]{CFXH19}, we can obtain better upper bounds on the code length based on binary constant weight codes. By Theorem \ref{thm1.1}, we can construct an $\ell\times (q^{2}+q+1)$-line-point incidence matrix $A$ whose rows correspond to the $\ell$ lines with Hamming weight $q+1$, and whose columns correspond to $q^{2}+q+1$ points in $PG(2,q)$.
Next, we give an upper bound on $\ell$ by the intersection condition.

\begin{lemma}\label{non-Sunflower}
Suppose $q\ge (\delta+1)$. If there exist $\ell$ distinct lines $L_{1},L_{2},\ldots,L_{\ell}$ in $PG(2,q)$  satisfying the intersection condition which do not form a Sunflower, then
\begin{equation*}
\ell\leq \left\lfloor (q^2+q+1)/(\delta+2)\right\rfloor.
\end{equation*}
\end{lemma}
\begin{IEEEproof}
Let $A$ be the $\ell\times (q^{2}+q+1)$-incidence matrix of the $\ell$ lines defined as above. The intersection condition implies that for each line $L_i$, there exists at least $(\delta+1)$ points on $L_i$ not belonging to any other lines, which implies that there exist at least $\ell(\delta+1)$ columns whose Hamming weight is 1. Equivalently, we assume that $A$ has the following structure:
\begin{equation}\label{6}
A=\left(\begin{array}{c|c}
  A'&
 \underbrace{ \begin{array}{cccc}
     \underbrace{  1 ~1 ~\cdots~ 1 }_{\delta+1}& 0 ~  0 ~\cdots~ 0 & \ldots & 0 ~ 0 ~\cdots~ 0 \\
     0 ~ 0 ~\cdots~ 0 &  \underbrace{  1 ~1 ~\cdots~ 1 }_{\delta+1} &\ldots & 0 ~ 0 ~\cdots~0 \\
   \vdots ~ \vdots ~ \vdots &  \vdots ~ \vdots ~ \vdots & \ddots & \vdots ~ \vdots ~ \vdots \\
 0 ~ 0 ~\cdots~ 0 & 0 ~ 0 ~\cdots~ 0 &\ldots & \underbrace{  1 ~1 ~\cdots~ 1 }_{\delta+1} \\
  \end{array}}_{\ell(\delta+1) ~\text{columns}}
  \end{array}
\right),
\end{equation}
where $A'$ is an $\ell \times (q^{2}+q+1-\ell(\delta+1))$ matrix. We can show that the Hamming weight of each column of $A'$ is at most $q-\delta$ since the $\ell$ lines do not form a sunflower and satisfy the intersection condition. Calculating the number of 1's in $A'$  in two ways, we obtain that $\ell \times (q-\delta)\leq (q^{2}+q+1-\ell(\delta+1))\times (q-\delta)$, i.e.,  $\ell\leq \left\lfloor (q^2+q+1)/(\delta+2)\right\rfloor.$
\end{IEEEproof}

From Theorem \ref{thm1.1} and Lemma \ref{non-Sunflower}, we have an improved upper bound as follows.
\begin{theorem} \label{thm5}
Suppose $q \geq \delta+1$ and let $C$ be a $q$-ary optimal $(n,k,d=2\delta+2;r=2,\delta)$ LRC with disjoint local repair groups, then
\begin{equation*}\label{7}
n\!\leq\! \max\left\{(\delta+1)(q+1), (\delta+1)\left\lfloor (q^2+q+1)/(\delta+2)\right\rfloor\right\}.
\end{equation*}
 \end{theorem}
 
By employing the Johnson upper bound (Lemma \ref{prop1}), we can obtain two improved upper bounds that are generalizations of Theorem 9 in \cite{CFXH19}.

\begin{theorem}
Suppose $q \geq \delta+2$ and a Singleton-optimal $(n,k,d=2\delta+2;r=2,\delta)$ LRC $\cC$ exists, then
\begin{eqnarray}\label{johnson1}
n \!\leq\! (\delta+1) \left\lfloor\frac{(2\delta+3)q^2+q+(\delta+1)^2-\sqrt{\Gamma}}{2(\delta+1)^2}\right\rfloor\!\!=\!O(q^2),
  \end{eqnarray}
where $\Gamma\triangleq(4\delta+5)q^4-(8\delta^2+12\delta+2)q^3+(4\delta^3+6\delta^2-1)q^2-2(\delta+1)^2q+(\delta+1)^4$. Moreover, we have
\begin{eqnarray}\label{johnson2}
n\!\!\!&\!\!\!\leq\!\!\!&\!\!\!(\delta+1)\left\lfloor\frac{(\delta(q+2)+2)+q\sqrt{4(\delta+1)q-(3\delta^2+4\delta)}}{2(\delta+1)}\right\rfloor\nonumber\\
&=&O(q^{1.5}) .
\end{eqnarray}
\end{theorem}

\begin{IEEEproof}
Let $A'$ be an $\ell \times (q^{2}+q+1-\ell(\delta+1))$-matrix defined in \eqref{6}. By Lemma \ref{non-Sunflower}, we know that the Hamming weight of each row of $A'$ is $q-\delta$. Since two lines in $PG(2,q)$ intersect at exactly one point, the Hamming distance of any two distinct rows of $A'$ is equal to $2(q-\delta)-2=2(q-\delta-1)$. Let $C'$ be a binary code whose codewords are the row vectors of $A'$, then $C'$ is exactly a binary $(n, M, d;w)$-constant weight code with $n=q^{2}+q+1-\ell(\delta+1)$, $M=\ell$, $d=2(q-\delta-1)$  and $w=q-\delta$. By plugging the parameters into the Johnson bounds (Lemma \ref{prop1}), we can obtain the desired upper bounds \eqref{johnson1}-\eqref{johnson2} on $n$ by solving two quadratic inequalities of $\ell$.
\end{IEEEproof}


\section{Conclusion}
In this paper, we provide a framework for further research on the maximum code length and constructions of optimal $(r,\delta)$ LRCs with small localities. Unlike known upper bounds that are asymptotically achieved, our improved upper bounds largely close the gap between the bounds and known construction when $2\delta+1\leq d\leq 3\delta$. It's worth noting that our bounds can be exactly achieved in some special cases. Furthermore, we establish a generalized characterization of optimal $(r=2,\delta)$-LRCs when $d=2\delta+2$. With this geometrical characterization, we obtain a sunflower based $(r,\delta)$ LRCs with longest code length $n=(\delta+1)(q+1)$ until now. 
 \begin{figure}[!htbp]
	\centering
	\includegraphics[width=0.9\linewidth]{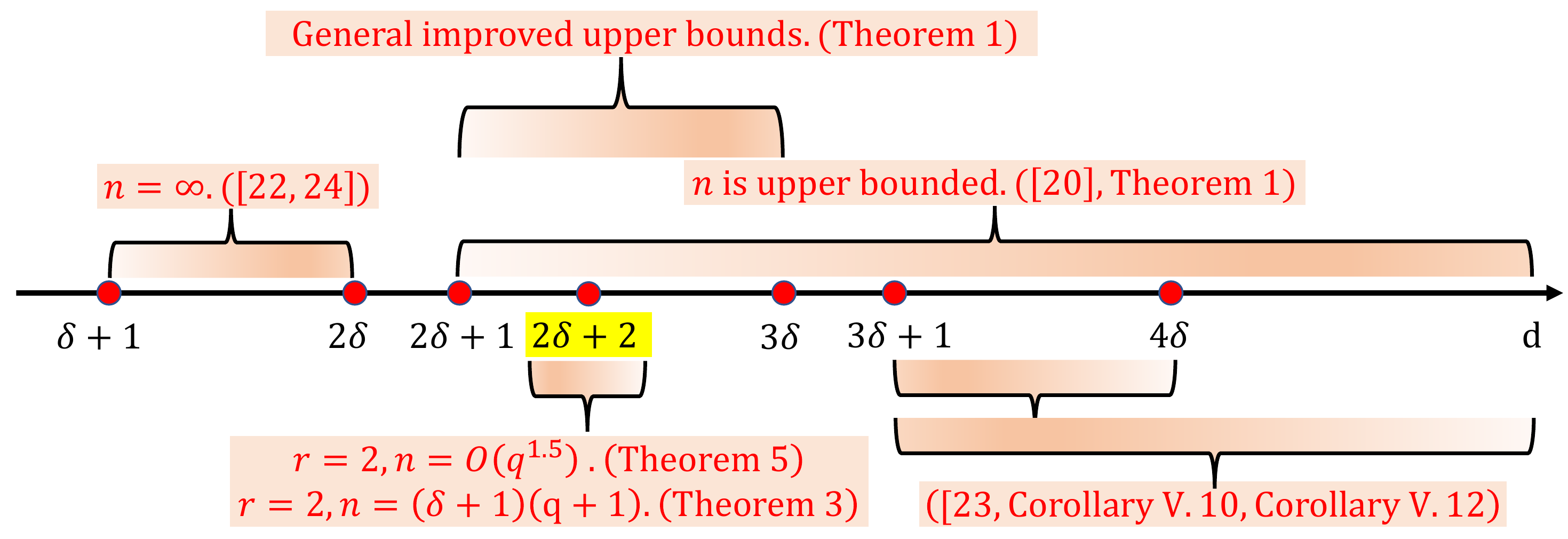}
	\caption{Summary of Related Progress.} 
	\label{fig1}
\end{figure}

We list known results in Figure \ref{fig1}. Despite these advances as shown in Figure \ref{fig1}, it is still a challenging problem to construct optimal $(r,\delta)$ LRCs achieving the maximal code length when $d\geq 2\delta+1$, especially when $\delta$ is odd that can not degenerate into the $r$-local case. Our future work will focus more on this parameter range, and generalize other combinatorial structures considered in \cite{CFXH19} to obtain more optimal $(r,\delta)$ LRCs with a longer length.

\bibliographystyle{IEEEbib}
\bibliography{refs}

\end{document}